# Phase-based stimulated emission depletion (pSTED) magnetic particle imaging

Guang Jia[1,*,#], Zhongwei Bian[2,*], Tianshu Li[3], Shi Bai[3,#], Yongchen Gou[1], Yiwen Li[1], Lixuan Zhao[4], Jia Luo[5], Mingli Peng[5], Weihua Li[6], Peng Gao[4], Tanping Li[4,#], Hui Hui[7], Jie Tian[7,#]

[1]School of Computer Science and Technology, Xidian University, Xi'an, Shaanxi, China;
[2]School of Biological Science and Medical Engineering, Beihang University, Beijing, China;
[3]School of Information Science and Engineering, Shenyang University of Technology, Shenyang, Liaoning, China;
[4]School of Physics, Xidian University, Xi'an, Shaanxi, China;
[5]College of Chemistry and Materials Science, Northwest University, Xi'an, Shaanxi, China;
[6]National Institute on Drug Dependence and Beijing Key Laboratory of Drug Dependence, Peking University, Beijing, China;
[7]CAS Key Laboratory of Molecular Imaging, Institute of Automation, Chinese Academy of Sciences, Beijing, China

*These authors equally contributed to this work.
#Corresponding author: Guang Jia (gjia@xidian.edu.cn), Shi Bai (baishi@sut.edu.cn), Tanping Li (tpli@xidian.edu.cn), Jie Tian (jie.tian@ia.ac.cn)

# Title Page

**Article type:** Research article

**Title:** Phase-based stimulated emission depletion (pSTED) magnetic particle imaging

**Running title:** pSTED MPI

## Abstract

Magnetic particle imaging (MPI) is an *in vivo* method to detect magnetic nanoparticles for cell tracking, vascular imaging, and molecular target imaging without ionizing radiation. Current magnetic particle imaging is accomplished by forming an field-free line (FFL) through a gradient selection field. By translating and rotating FFL under excitation and drive fields, the harmonic complex signal of a point source forms a Lorentzian-shape point spread function on the plane perpendicular to FFL. The Lorentzian PSF has a finite size and limited resolution due to the non-sharp Langevin function and weak selection field. This study proposes a donut-shaped focal spot by borrowing the stimulated emission depletion (STED) fluorescence microscopy principle. The influence of the gradient selection field on the relaxation time of magnetic particles determines the nonlinear phase shift of the harmonic complex signals, resulting in the formation of a donut-shaped focal spot. By subtracting the donut-shaped focal spot from the Lorentzian focal spot, the STED focal spot size was reduced by up to 4 times beyond the Langevin magnetization resolution barrier. In human brain FFL-based MPI scanner, the donut-shaped focal spot can be used to reconstruct images with super-resolution and super-sensitivity through the deconvoution of the STED focal spot and filtered backprojection algorithm.

**Introduction**

Magnetic particle imaging (MPI) is an *in vivo* method without ionizing radiation for cell tracking[1-3], vascular imging[4-6], and molecular target imaging[7-10]. MPI was conceptually proposed in 2001 and experimentally proved in 2005[11]. Recently developed MPI scanners used the selection field to generate an field-free line (FFL) for signal collection from magnetic particles along the FFL[12,13]. The parallel translation of FFLs is needed to acquire a projected image of magnetic particles[14,15], in addition to which the rotations of FFLs and filtered backprojection algorithm are required to generate a cross-sectional image of magnetic particles[16,17]. This type of MPI scanner design was constructed and validated for human brain imaging[18]. The image resolution of large-bore or brain-sized MPI devices is about 1-10 mm[5,18-20], which is resulted from the non-sharp derivation of the Langevin magnetization function and a weak selection field gradient[15,21].

A donut-shaped focal spot based on the stimulated emission depletion (STED) method has been successfuly used to improve the image resolution of fluorescence microscopy. Professor Stefan Hell proposed the central-dark donut focal spot[22] and achieved a 3-fold resolution improvement[23], which spans from the "micron" microscopy era to the "nano" imaging era[24]. Professor Hell's lab recently proposed the MINSTED technology[25], by changing the radius and position of the donut focal spot, further accurately locating the fluorescent spots and improving the resolution[26].

A donut-shaped focal spot from the STED fluorescence imaging principle may be introduced into MPI to enhance the imaging resolution. In an FFL-based MPI with collinear excitation and receiver parallel to the FFL, the projected harmonic signal map of a point source phantom perpendicular to the FFL is shown as a Lorentzian-shaped focal spot[15,27,28]. The relaxation time of magnetic nanoparticles decreases with the increase of selection gradient stimulation field[29] and causes the phase lag of the harmonic complex signal[30]. The donut-shaped focal spot can be generated by rotating the harmonic complex signal to null the imaginary part of the harmonic signal at the center of the FFL. The subtraction of the donut focal spot from the Lorentzian focal spot provides a significantly small PSF to enhance the MPI resolution and sensitivity.

## STED-MPI Theory

The selected gradient field of an FFL-based MPI scanner is as follows:

$$H_{SL}(x,y,z) = \begin{pmatrix} G & 0 & 0 \\ 0 & G & 0 \\ 0 & 0 & 0 \end{pmatrix} \cdot \begin{pmatrix} x \\ y \\ z \end{pmatrix} = \begin{pmatrix} Gx \\ Gy \\ 0 \end{pmatrix}, \tag{1}$$

in which $G$ is the magnetic field gradient in both $x$ and $y$ directions in Fig. 1a.

If the alternating excitation field along $z$ direction has a frequency $\omega_0 = 2\pi f/=2\pi/T$ and an amplitude of $H_{AC}$.

$$H_{EX}(t) = \begin{pmatrix} 0 \\ 0 \\ -H_{AC}cos\omega_0 t \end{pmatrix}. \tag{2}$$

Using a pair of receiver coils along the $z$ direction (collinear excitation and receiver setting), the signal is in a convolution form,

$$v(t) = v_{ad}(t) * r_D(t), \tag{3}$$

where $r_D(t)$ is the Debye relaxation kernel of magnetic nanoparticles, defined as follows:

$$r_D(t) = \frac{1}{\tau} e^{-t/\tau} u(t). \tag{4}$$

where $u(t)$ is the Heaviside step function, and $\tau$ is the relaxation time. With Fourier transform, the $n$th harmonic of signal frequency components is defined as

$$V_n(r) = V_{n,ad}(r) \cdot R_n(r), \tag{5}$$

where the relaxation frequency components are defined as

$$R_n(r) = \frac{\omega_0}{2\pi} \frac{e^{-i\theta_{n,\tau(r)}}}{\sqrt{1+(n\omega_0\tau(r))^2}}, \tag{6}$$

where $\theta_{n,\tau(r)}$ is delay phase due to the relaxation effect (Fig. 1b),

$$sin(\theta_{n,\tau(r)}) = \frac{n\omega_0\tau(r)}{\sqrt{1+(n\omega_0\tau(r))^2}} \text{ and } cos(\theta_{n,\tau(r)}) = \frac{1}{\sqrt{1+(n\omega_0\tau(r))^2}}. \tag{7}$$

The relaxation time is dependent upon the stimulation field along $z$ direction and gradient field along $r$ direction[29,31,32],

$$\tau(r) = \frac{2L(\xi)}{\xi - L(\xi)}\tau(0), \tag{8}$$

where $\xi = \beta|H| = \frac{\mu_0 m}{k_B T^P}\sqrt{(Gx)^2 + (Gy)^2 + H_{AC}^2}$ and $\tau(0) = \frac{1}{\sqrt{1+0.126(\beta H_{AC})^{1.72}}}\frac{3\eta V_h}{k_B T^P}$.

We simply assume a 1-dimensional direction signal harmonic $V_n(r)$ as the magnetic nanoparticle concentration $c(r)$ and the point spread function $G_n(r)$,

$$V_n(r) = c(r) * G_n(r). \tag{3}$$

The $n$th harmonic signal PSF $G_n(r)$ along the the $r$ axis can be calculated as

$$G_n(r) = G_{n,ad}(r)\frac{\omega_0}{2\pi}\frac{e^{-i\theta_{n,\tau(r)}}}{\sqrt{1+(n\omega_0\tau(r))^2}}, \tag{9}$$

in which the adiabatic delta response of the system $G_{n,ad}(r)$ is in a Lorentzian form[15,28] with two fitted parameters ($\alpha_n$ and $\beta_n$),

$$G_{n,ad}(r) = \beta_n \cdot \frac{\alpha_n}{r^2+\alpha_n^2}. \tag{10}$$

Through phase modulation, it is possible to clock out the $n$th harmonic PSF $G_n(r)$ by an angle of $\theta_{n,\tau(0)}$, such that the imaginary part is zero at $r = 0$ (Fig. 1c), and becomes

$$G_n(r) = G_{n,ad}(r)\frac{\omega_0}{2\pi}\frac{e^{i(\theta_{n,\tau(0)}-\theta_{n,\tau(r)})}}{\sqrt{1+(n\omega_0\tau(r))^2}}, \tag{11}$$

in which the real part $G_{nR}(r)$ is a regular focal spot,

$$G_{nR}(r) = G_{n,ad}(r)\frac{\omega_0}{2\pi}\frac{1+(n\omega_0)^2\tau(r)\tau(0)}{(1+(n\omega_0\tau(r))^2)\sqrt{1+(n\omega_0\tau(0))^2}}, \tag{12}$$

and the imaginary part $G_{nI}(r)$ is a donut-shaped focal spot,

$$G_{nI}(r) = G_{n,ad}(r)\frac{\omega_0}{2\pi}\frac{n\omega_0(\tau(0)-\tau(r))}{(1+(n\omega_0\tau(r))^2)\sqrt{1+(n\omega_0\tau(0))^2}}. \tag{13}$$

The relationship between the real and imaginary parts can be defined as an STED curve, which is expressed as a function of the relaxation time $\tau(r)$,

$$G_{nI}(r) = G_{nR}(r)\frac{n\omega_0(\tau(0)-\tau(r))}{1+(n\omega_0)^2\tau(r)\tau(0)}. \tag{14}$$

We define an STED factor $\alpha_{STED}$ to obtain the STED-based $n$th harmonic PSF (Fig. 1d),

$$G_{nSTED}(r) = G_{nR}(r) - \alpha_{STED} \cdot G_{nI}(r). \tag{15}$$

The full-width half maximum (FWHM) of the $n$th harmonic PSF is a function of the STED factor. With the optimal STED factor, we can achieve 2 to 3 fold increase in the resolution with the minimal negative subtracted signals.

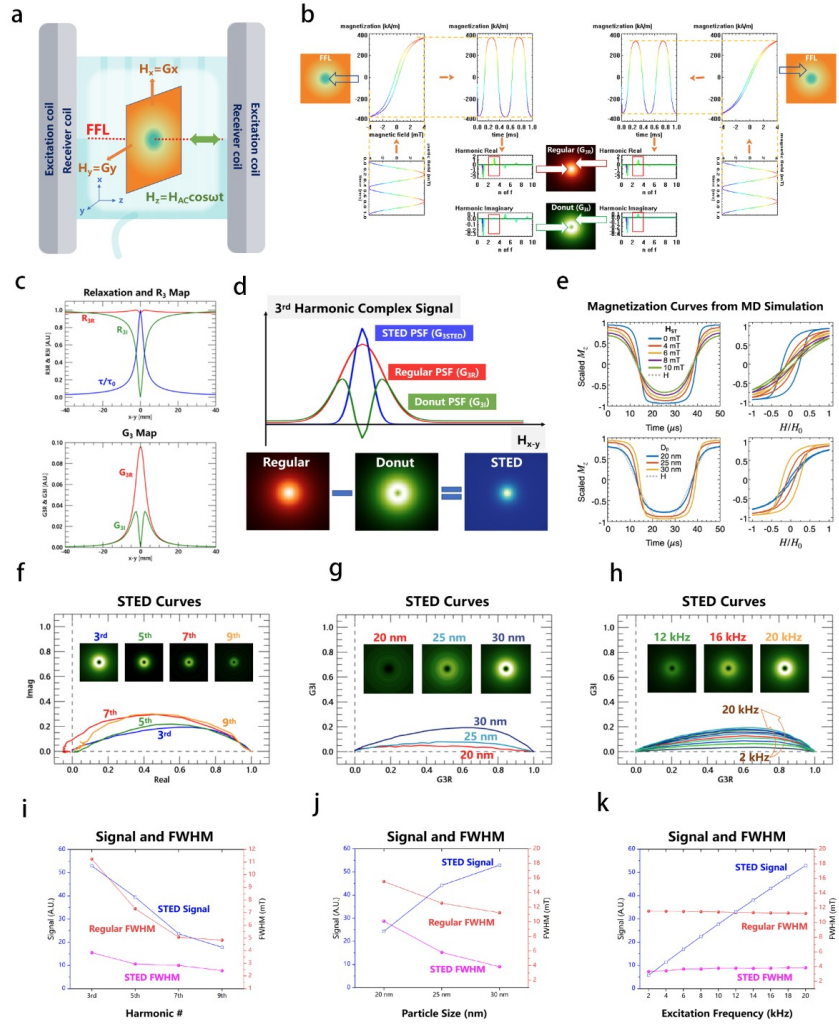

**Figure 1 | STED-MPI scanner concept and molecular dynamics simulation of Donut-shaped PSF. a,** STED-MPI is based on a collinear setting with parallel FFL, excitation, and receiver. The selection gradient field perpendicular to the FFL behaves as a stimulation field resulting in signal depletion. **b,** Magnetic nanoparticles at the center or vicinity of the FFL have different magnetization curves. The imaginary part of the 3rd harmonic signal can be zero at the FFL center and displayed as a donut-shaped PSF on the *x-y* plane. The real part of the 3rd harmonic signal can be Lorentzian shape. **c,** The relaxation time decreases with the increasing stimulation field and the imaginary parto of the Debye relaxation kernel can be zero at the FFL center, resulting a smaller donut-shaped PSF. **d,** The subtraction of the donut PSF (3rd harmonic imaginary signal) from the regular PSF (3rd haronic real signal) generates a much smaller PSF, which is the so-called STED PSF for high resolution MPI. **e,** The magnetization curves are different due to the relaxation time at different stimulation field or with different particle size. **f,** The curve of the real and imaginary part of the harmonics versus the stimulation field is defined as the STED cuve. The higher harmonic STED curve generates smaller donut-shaped PSF but a lower signal, as shownn in **i**. **g,** Magnetic nanoparticles with larger size exhibits a greater signal and smaller donut PSF with smaller FWHM, as shown in **j**. **h,** The higher excitation frequency does not affect the FWHM or image resolution, but increasing the signal, as shown in (**k**).

**FFL-based MPI Scan Experiment**

In recently developed FFL-based MPI scanner, there is only 2D images by using the translation of an FFL on the x-y plane (Fig. 2a). The 2-dimensional 3${}^{rd}$ harmonic signal map $V_3(\vec{r})$ as the magnetic nanoparticle concentration $c(\vec{r})$ and the point spread function $G_3(r)$,

$$V_3(r) = c(r) * G_3(r), \tag{16}$$

which is a complex value and can be separted into a real and imaginary part,

$$V_{3R}(r) = c(r) * G_{3R}(r) \text{ and } V_{3I}(r) = c(r) * G_{3I}(r) \tag{17}$$

By using an STED factor, the harmonic signal can be

$$V_{3STED}(r) = c(r) * G_{3STED}(r). \tag{18}$$

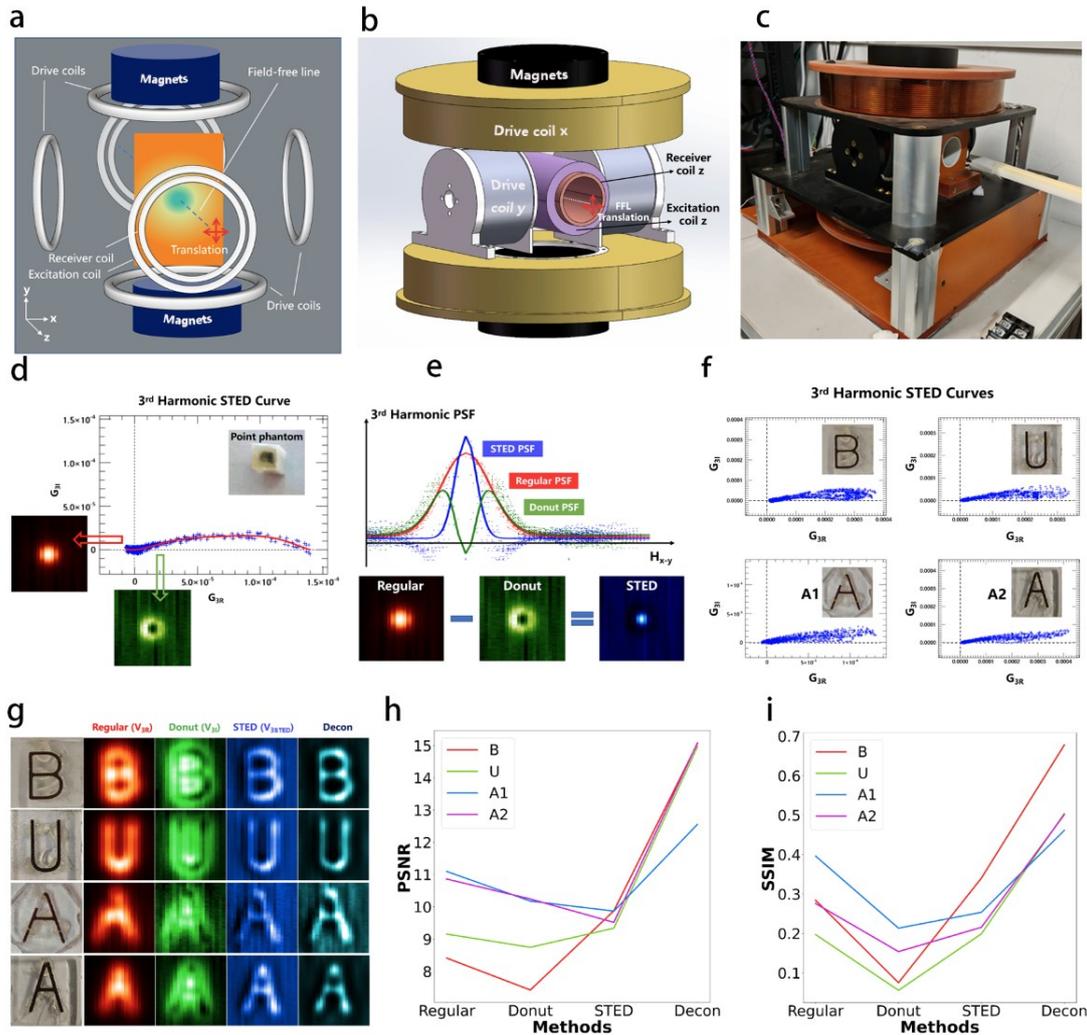

Figure 2 | STED-MPI scanner with FFL translations and reconstructed images. **a,** The main components of a STED-MPI scanner that requires collinear FFL, excitation and receiver. **b,** Drive coils push the parallel movement of the FFL and generate projeced 2D images. **c,** The picture of the STED-MPI scanner. **d,** The 3$^{rd}$ harmonic STED curve of a point source phantom. The real part can be a regular Lorentzian curve and the imaginary part a donut-shaped PSF after centain angle rotation. **e,** A 3$^{rd}$ harmonic STED PSF can be derived by combining the regular PSF and donut-shaped PSF. **f,** The 3$^{rd}$ harmonic curves of letter phantoms. The complex 3$^{rd}$ harmonic components were rotated by using the same angle that generates the donut-shaped PSF. **g,** The reconstructed images of the letter phantoms by using regular PSF, donut PSF, STED PSF, and STED deconvolution methods. The STED deconvolution method shows highest peak signal-to-noise ratio (PSNR) in **h** and greatest he structural similarity index measure (SSIM) in **i**.

## FFL-based MPI Scan Simulation

In recently developed human brain FFL-based MPI scanners[18], FFL is formed along the $y'$-axis and rotated along the $z$-axis to obtain MNP concentration projection signals at different angles, forming a projection sinogram (Fig. 3a). The selection gradient field of the FFL along $y'$ direction is defined as

$$H_{SL}(x',y',z) = \begin{pmatrix} G & 0 & 0 \\ 0 & 0 & 0 \\ 0 & 0 & G \end{pmatrix} \cdot \begin{pmatrix} x' \\ y' \\ z \end{pmatrix} = \begin{pmatrix} Gx' \\ 0 \\ Gz \end{pmatrix}, \qquad (19)$$

in which $G$ is the magnetic field gradient in both $x'$ and $z$ directions. The alternating excitation field along $z$ direction is

$$H_{EX}(t) = \begin{pmatrix} 0 \\ 0 \\ -H_{AC}\cos\omega_0 t \end{pmatrix}. \qquad (20)$$

Using a pair of receiver coils along the $z$ direction (collinear excitation and receiver setting), the projected 3$^{rd}$ harmonic signal along the axis $x'$ from an angle of $\theta$ is $V_3(x',\theta)$. The real and imaginary signal components, $V_{3R}(x',\theta)$ and $V_{3I}(x',\theta)$, are based on Lorentzian and donut PSF respectively. An STED signal part can be generated,

$$V_{3STED}(x',\theta) = V_{3R}(x',\theta) - \alpha_{STED} \cdot V_{3I}(x',\theta), \qquad (21)$$

where the STED factor, $\alpha_{STED}$, was determined by the STED-MPI theory based on the relaxation property of magnetic nanoparticles and scan parameters. Fig. 3b shows the approximate projection curves of Regular ($V_{3R}$), Dount ($V_{3I}$), and STED ($V_{3STED}$) of the scanned phantom slice at different projection angles during the imaging process.

The STED signals with multiple projection angles form into a sinogram, which can be used to reconstructed MNP concentration maps via the deconvolution process and filtered backprojection algorithm[28],

$$c(x,y) = \mathcal{R}^{-1}\{V_{3STED}(x',\theta)/G_{3STED}(x')\}, \qquad (22)$$

where $\mathcal{R}^{-1}$ the inverse Radon transform operator[33]. Under the condition of 27 projections and 132 individual projection points, the obtained Regular ($V_{3R}$), Dount ($V_{3I}$), and STED ($V_{3STED}$) sinogram and the reconstruction results obtained by filtered back projection on the sinogram are shown in Fig. 3c. It can be observed that the STED method has clearer edges and higher imaging quality compared to the Regular method for digital phantom reconstruction, reflecting the superiority of the proposed algorithm STED in simulating imaging processes. During the simulation process, we observed aliasing artifacts on the reconstructed images, which is due to the fact that the number of projections was smaller than the theoretical minimum projection number obtained from the formula for eliminating artifacts[28].

In order to investigate the influence of projection quantity on the reconstruction results, we tested the Regular and STED image reconstruction results under different projection quantity conditions, and plotted the variation curves of peak signal-to-noise ratio (PSNR) and structural similarity (SSIM) of the reconstruction results with increasing projection quantity, as shown in Fig. 3d and 3g. As the number of projections gradually increases, the PSNR and SSIM indicators of the reconstructed results gradually increase and eventually tend towards constant values. In addition to the number of projections, the number of segments in a single projection is also an important factor affecting the quality of the reconstruction results.

Fig. 3e and 3h show the reconstruction results of the Regular method and STED method at different number of segments per projection, as well as the PSNR and SSIM indicators of the reconstruction results changing with the number of projection points. As the number of segments per projection changes, PSNR and SSIM show significant fluctuations, and as the number of segments per projection increases, the fluctuations of PSNR and SSIM gradually decrease, and the magnitude of the increase gradually decreases.

Fig. 3f and 3i show the results of reconstruction using four commonly used filtering kernels in filtered back projection under simulation conditions with 100 projections and 237 segments per projection. It can be observed that the reconstruction results of different filtering kernels are relatively similar, indicating that the impact of different commonly used filtering kernels on the results is not significant when performing FBP under the Regular method and STED method.

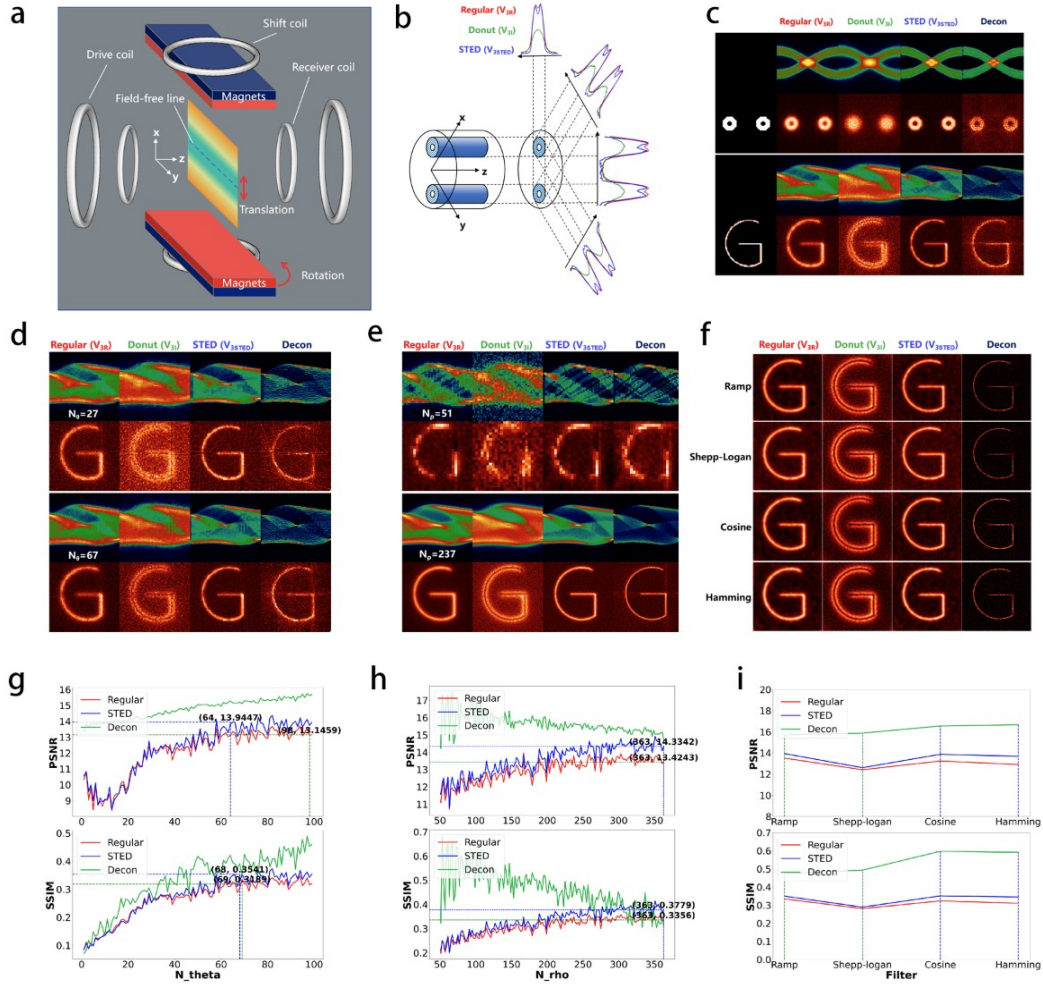

Figure 3 | STED-MPI with FFL rotation and reconstructed images. **a,** The main components of a human brain MPI scanner with parallel excitation and receiver, and perpendicular FFL. The translation and rotation of FFL generate 2D images on *x-y* plane. **b,** The selection gradient field perpendicular to FFL (*y*-direction), excitation (*z*-direction), and receiver (*z*-direction) can generate regular Lorentzian PSF or donut shaped PSF, the subtraction of which can result in a smaller STED PSF. **c,** Sinograms and reconstructed images of hollow plates and 'G'-shaped phantoms using filtered backprojection algorithm based on regular PSF, donut PSF, STED PSF, and STED deconvolution. **d,** Sinograms and reconstructed images of "G"-shaped phantom by using different number of projection angles. The PSNR and SSIM increase with a increasing number of projection angles in **g**. STED deconvolution method exhibits a significant higher PSNR and SSIM than the other methods. **e,** Sinograms and reconstructed images of "G"-shaped phantom by using different number of segments per projection. The PSNR and SSIM of STED deconvolution method are higher than the other method and decrease with a greater number of segments per projection in **h**. **f,** Reconstructed images based on different filtering kernels. The PSNR and SSIM of STED deconvolution method are higher than the other method (**i**). The filtering kernels show little effect on reconstructed image quality.